\documentclass[sigconf]{acmart}

\usepackage{algorithm2e}

\begin{document}

\title{SQLFlow: A Bridge between SQL and Machine Learning}

\author{Yi Wang}
\author{Yang Yang}
\author{Wei Yan}
\author{Mingjie Tang}
\author{Yuan Tang}
\email{yi.w,yang.y,wei.yan,}
\email{m.tang,yuan.tang@antfin.com}
\affiliation{%
  \institution{Ant Financial USA}
  \city{Sunnyvale}
  \state{California}
}

\author{Weiguo Zhu}
\author{Yi Wu}
\author{Xu Yan}
\email{weiguo.zhuwg,xiongmu.wy,yancey.xy,}
\author{Yongfeng Liu}
\author{Yu Wang}
\email{antonov.lyf,wy232418@antfin.com}
\affiliation{%
  \institution{Ant Financial}
  \city{Hangzhou}
  \country{China}
}

\author{Liang Xie}
\author{Ziyao Gao}
\author{Wenjing Zhu}
\author{Xiang Chen}
\email{xieliang,gaoziyao,zhuwenjing,}
\email{alfredchenxiang@didiglobal.com}
\affiliation{%
  \institution{DiDi}
  \city{Beijing}
  \country{China}
}

\begin{abstract}
Industrial AI systems are mostly end-to-end machine learning (ML) workflows. A typical recommendation or business intelligence system includes many online micro-services and offline jobs. We describe SQLFlow for developing such workflows efficiently in SQL.  SQL enables developers to write short programs focusing on the purpose (what) and ignoring the procedure (how).  Previous database systems extended their SQL dialect to support ML. SQLFlow\footnote{SQLFlow is an open-source project at \url{https://sqlflow.org/sqlflow}} takes another strategy to work as a bridge over various database systems, including MySQL, Apache Hive, and Alibaba MaxCompute, and ML engines like TensorFlow, XGBoost, and scikit-learn. We extended SQL syntax carefully to make the extension working with various SQL dialects.  We implement the extension by inventing a collaborative parsing algorithm.  SQLFlow is efficient and expressive to a wide variety of ML techniques -- supervised and unsupervised learning; deep networks and tree models; visual model explanation in addition to training and prediction; data processing and feature extraction in addition to ML. SQLFlow compiles a SQL program into a Kubernetes-native workflow for fault-tolerable execution and on-cloud deployment. Current industrial users include Ant Financial, DiDi, and Alibaba Group.
\end{abstract}

\maketitle

\section{Introduction}

\begin{figure}[b]
  \centering
  \includegraphics[width=\linewidth]{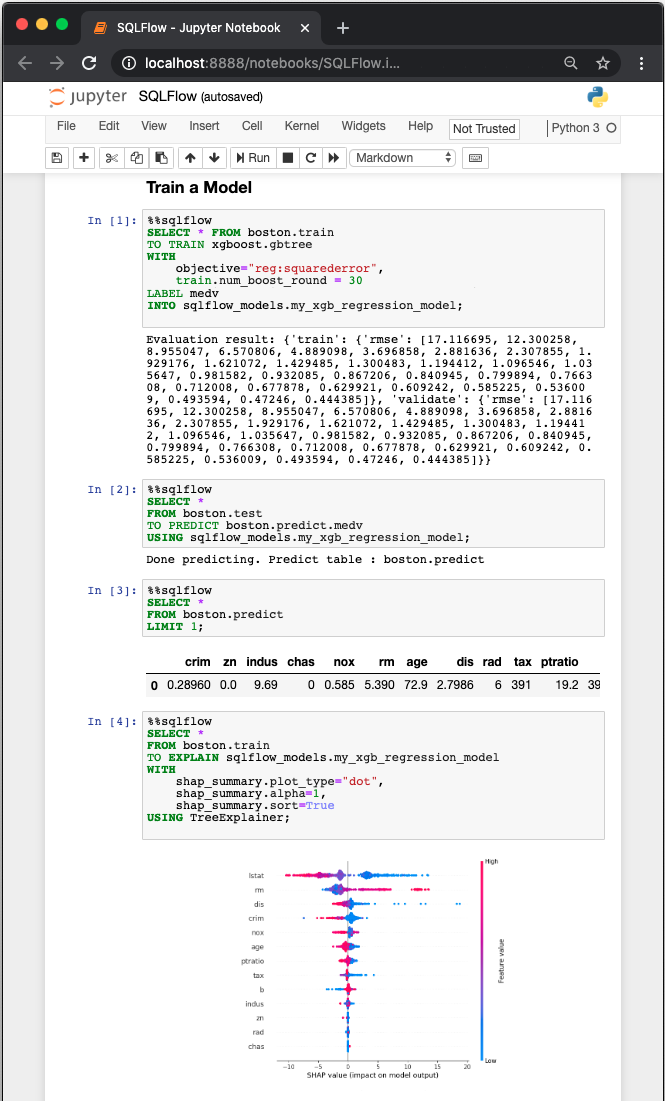}
  \caption{SQLFlow with Jupyter Notebook as the GUI.}
  \Description{SQLFlow with Jupyter}
\end{figure}

Most industrial AI systems are end-to-end machine learning workflows of multiple steps. Typical workflows include search engines, recommendation systems for e-commerce, and business analysis platforms like Salesforce.  A workflow usually starts with a Web server that creates the UI, followed by log collectors like fluientd, log streams aggregators like Kafka, database systems like Apache Hive, and ML systems built on top of TensorFlow~\cite{tensorflow}, PyTorch~\cite{pytorch}, XGBoost~\cite{xgboost}, or scikit-learn~\cite{scikit-learn}. Most workflows are cyclic because the trained models help the Web server to answer more user questions via an online serving system or offline scoring system.  Each of such a workflow might need tens or hundreds of engineers to build and continuously improve.

A recent trend to simplify the development of end-to-end ML workflows, with the commonly used components getting modulized, is to program and deploy the workflow as a whole using expressive languages like Python. Examples include Amazon SageMaker~\cite{sagemaker}, Netflix Metaflow~\cite{metaflow}, and BlazingSQL~\cite{blazingsql}.

We propose SQLFlow to program in SQL other than Python. SQL is unique as it allows programmers to focus on their purposes while ignoring the procedure of achieving that purpose, thus improves the development efficiency significantly.  Figure~1 shows how to train an XGBoost model, make a prediction, and visually explain the model by calling SHAP~\cite{shap}.  The SQLFlow syntax extension used in this SQL program takes care of data pre-processing and extracts features automatically.

Some database systems extend their SQL dialects to support ML.  For example, Google BigQuery~\cite{bigquery}, a database system running on Google Cloud, adds the CREATE MODEL statement for training ML models.  Other examples include Microsoft SQL Server Machine Learning Services~\cite{sqlserver}, Azure SQL Database~\cite{azure}, Amazon Athena~\cite{athena}, IBM DB2~\cite{db2}, Apache Madlib~\cite{madlib}, and MemSQL~\cite{memsql}.  However, these approaches have some drawbacks.

\begin{itemize}
\item They lock end-to-end ML users to specific database systems.  Switching to another database system requires rewriting of all ML solutions.
\item Most of the above systems and their extensions are not open-source, thus prohibits community contributions of ML features.
\item These solutions are based on existing database computing infrastructures, and cannot use the fault-tolerance and extensibility of general cluster computing technologies like Kubernetes.
\end{itemize}

SQLFlow takes a different strategy.  It is open-sourced and works like a bridge connecting various database systems and ML engines.  Since open-sourced in May 2019, the community added support to MySQL, Apache Hive, Alibaba MaxCompute, TensorFlow, XGBoost, scikit-learn, and SHAP.  The list is growing.  To make the ML syntax extension to work with various SQL dialects, we carefully designed the extension and invented a collaborative parsing algorithm, which calls parsers of SQL dialects and that of the syntax extension.

Instead of running ML workloads on database infrastructure, SQLFlow compiles SQL programs into workflows for deployment and execution on Kubernetes, which is overwhelming public clouds and on-premise clusters.  Workflow execution engines, like Argo\footnote{https://argoproj.github.io} and Tekton\footnote{https://github.com/tektoncd/pipeline}, run on Kubernetes and provide fault-tolerance and scalability.  SQLFlow-generated workflows access the database systems and ML engines, which also run on Kubernetes.  The compilation algorithm automatically generates workflow steps like data cleanup and feature extraction to fill the gap between raw data and ML.

In the ecosystem of end-to-end ML, there are three roles:
\begin{itemize}
    \item End-to-end ML users call ML models from SQL.
    \item ML researchers define models in Python and abstraction APIs like Keras.
    \item Tool developers create ML engines and database systems in mostly C++ and Go.
\end{itemize}
ML researchers need to write workflows to verify and test model definitions.  When doing so, they prefer using their well-used tool Python.  However, Argo and Tekton accept YAML files, which are lengthy and error-prone to write.  To flourishing the community, SQLFlow provides Couler, a Python library for describing workflows.  Tool developers can contribute Couler functions to encapsulate their tools, for example, \verb|couler.train_xgboost_model| and \verb|couler.run_mysql_query|.  ML researchers can call these functions to build up their workflows.  SQLFlow translates SQL programs in Couler programs to ease of debugging.  Couler then translates Python programs into YAML files.

The open architecture of SQLFlow poses some restrictions. For example, we don't yet have a plan to support transactions shortly. Users can use the transaction mechanism provided by the database systems in their SQL programs, as the generated workflow submits "normal SQL statements" to the database system for execution. However, SQL statements with SQLFlow syntax extensions can not be in part of the transactions.

\section{Syntax Extension}

Some prior work adds user-defined functions (UDFs) to a database system to support ML. For SQLFlow, however, it is intractable to implement the UDFs for all combinations of the supported database and ML systems.  Some other works introduce new statements.  For example, Google BigQuery has the CREATE MODEL statement.  We noticed that many data analysts have SELECT statements for data cleaning, data transformation, and feature extraction.  The next thing in their mind is naturally ML.  It is intuitive to append a clause for the purpose. SQLFlow introduces the following clause extensions.

\begin{verbatim}
    SELECT ... TO TRAIN model_definition 
                    WITH parameters 
                    COLUMN features LABEL target 
                    INTO model_name;
    SELECT ... TO PREDICT field USING model_name;
    SELECT ... TO EXPLAIN model_name WITH parameters;
\end{verbatim}

The TRAIN clause trains a model using the result from the SELECT statement, where model\_definition names a Python class that defines a model.  For example, DNNRegressor, or prefixed with the package name, dnn.Regressor, and even Docker image name, sqlflow/models.dnn.Regressor.  The identifier model\_name names an output folder for saving the trained model parameters.  It can be a URL pointing to external storage like Amazon S3.  Users can specify ML parameters using the WITH clause.  For example, WITH learning\_rate=0.01, hidden\_units=[10, 100, 15].  We are working on automatic parameters tuning using Katib\footnote{\url{https://github.com/kubeflow/katib}}.   The COLUMN and LABEL clauses specify the input and output of models.  LABEL is not required when training unsupervised models.

The PREDICT clause makes a prediction using a trained model.  The EXPLAIN clause visually explains a model.

The word TO before TRAIN, PREDICT, and EXPLAIN is necessary for disambiguation.  In some SQL dialects, for example, MySQL, the SELECT statement could have aliases.  Users can write SELECT * FROM t1, t2, t3 TRAIN ... where TRAIN is an alias of table t3.  As a reserved keyword in the SQL language specification, various dialects don't accept TO as table or alias name.

\section{Collaborative Parsing}

The collaborative parsing algorithm allows SQLFlow to understand ``normal statements'' of SQL dialects, and those SELECT statements followed by SQLFlow extension clauses.  This parsing algorithm can use any open-source parser, like the MySQL/TiDB, HiveQL, and Calcite parser, and proprietary parser under grants when we integrate with non-open-source systems.  This parsing algorithm also calls the SQLFlow extension parser written in goyacc and understanding only the TO TRAIN, PREDICT, and EXPLAIN clauses.  As shown below, the collaborative parser repeatedly calls the dialect parser via parseDialect and the SQLFlow extension parser via parseExtension.

\begin{algorithm}
\DontPrintSemicolon 
\KwIn{A SQL program with optional SQLFlow extensions}
\KwOut{Parsed statement list and optionally an error}
lst $\leftarrow$ []\;
\While{True}{
    stmts, stop\_at, err $\leftarrow$ parseDialect(program)\;
    \uIf{err $\neq$ nil}{
        \Return [], err\;
    }
    lst $\leftarrow$ append(lst, stmts)\;
    \uIf{stop\_at the end of program}{
        return lst, nil
    }
    program $\leftarrow$ program[stop\_at:]\;
    \;
    ext, stop\_at, err $\leftarrow$ parseExtension(program)\;
    \uIf{err $\neq$ nil}{
        \Return [], err\;
    }
    lst $\leftarrow$ appendExtension(lst, ext)\;
    \uIf{stop\_at the end of program}{
        \Return lst, nil
    }
    program $\leftarrow$ program[stop\_at:]\;
}
\end{algorithm}

The function parseDialect calls the dialect parser up to twice.  The first call would fail if there is an SQLFlow extension clause in the program and stop at the beginning of the clause.  In this case, parseDialect makes a second call to parse the part of the SQL program before the error position, which is before the extension clause.  If there are no syntax errors in that part, the second call succeeds.  This algorithm relies on the convention that the dialect parsers report error position, which is true because parsers are supposed to report errors position. They do this in different ways; for example, the MySQL/TiDB parser returns strings, but HiveQL and Calcite parser throw exceptions. We have a thin layer of abstraction to shield these minor differences.

\section{Compile SQL to Workflow}
\label{sec:compile}

SQLFlow can run as a gRPC server or a command-line tool.  In either way, it compiles a SQL program into a workflow.  We used to represent a workflow by a Python program, which, however, is not fault-tolerable.  In industrial deployments, like on Kubernetes clusters, the SQLFlow service runs as a group of replicated processes, and Kubernetes might preempt some processes when scheduling.  If the workflow runs as a Python program, they are subject to preemption.  As a solution, we changed SQLFlow to generate workflows in the form of YAML files for the execution by fault-tolerable engines like Argo and Tekton.

\subsection{Python API of Workflow}

The Argo/Tekton YAML files are hard to read and debug.  So we wrote a Python library Couler to represent workflows as Python source code.  The following 6-line Python program runs and prints a 46-line YAML file representing an equivalent workflow task of two steps.  The YAML file are for execution by Argo/Tekton.  In practice, we have Couler functions corresponding to ML and database systems like \verb|train_xgboost_model| and \verb|run_mysql_query|.

\begin{verbatim}
    @couler.task
    def echo_hello_world(hello, world="El mundo"):
        couler.step(cmd=["echo"], args=[hello])
        couler.step(cmd=["echo"], args=[world])
    echo_hello_world("Aloha")    
\end{verbatim}

\subsection{Two-Tier Compilation}

SQLFlow features a two-tier compilation architecture.  The top one translates a SQL program into a YAML, where each SQL statement is a step.  Argo and Tekton run each step as a container, inside which, SQLFlow translates the SQL statement into a Python program and runs it.  If the statement is a normal one, the generated Python program submits it literally to the database system for execution; or, if it is a SELECT statement with SQLFlow extension, the Python program retrieves data, extracts features, and launches the ML task.

The two-tier compilation is necessary because SQLFlow needs to do feature extraction, an essential part of end-to-end machine learning, at run-time, after the execution of previous SQL statements and the schema of the input data table is known.  The second tier compilation happens at the run-time of the first tier.

\section{Demo Overview}

\begin{table*}
  \caption{Episodes in the Demo}
  \label{tab:demo}
  \begin{tabular}{lllllll}
    \toprule
      Episode  & Database system    & Database deployment   & ML system     & Model explanation & Learning paradigm  & UI \\
  \midrule
One.   & MySQL              & local        & XGBoost       & SHAP              & Supervised         & Jupyter \\
Two.   & Apache Hive        & Docker       & TensorFlow    & SHAP              & Supervised         & Jupyter \\
Three. & Alibaba MaxCompute & Cloud        & scikit-learn  & metaplot          & Unsupervised       & Jupyter \\
Four.  & MySQL              & Kubernetes   & XGBoost       & a text-backend for SHAP  & Supervised         & Terminal \\
  \bottomrule
  \end{tabular}
\end{table*}

The objectives of this demo are to show SIGMOD attendees: (1) the superior efficiency of end-to-end machine learning experience provided by an open architecture that connects various database and ML systems, (2) the expressiveness of SQLFlow syntax extension on both supervised and unsupervised learning, and (3) the versatility to integrate with GUI systems and to work alone as command-line compiler toolchain.

We plan to run SQLFlow as a local service so that users can interact with it via Jupyter Notebook or from the command-line. We plan to preload datasets to MySQL and Hive running locally and to MaxCompute running on the public cloud; users can access them from SQLFlow.  We provide a supplementary video with this paper showing four episodes; each is a typical use case.  As shown in Table~\ref{tab:demo}, these examples cover four datasets, various combinations of database and ML systems, tree models and deep learning models, supervised and unsupervised learning.  They also cover visual model explanations in the GUI and a text-based terminal. The audience can freely change example SQL programs used in the episodes and invent their own. They are free to invent new combinations out of Table~\ref{tab:demo}, like unsupervised learning with MySQL and XGBoost models with Hive.

\paragraph{Episode One.}  This episode shows SQLFlow works with MySQL and XGBoost on supervised learning of tree models.  It also shows the visual model explanation in Jupyter Notebook by calling SHAP.  The dataset is Boston Housing\footnote{\url{https://www.cs.toronto.edu/~delve/data/boston/bostonDetail.html}}. This episode includes the following steps. (1) Dataset examination by SELECT few rows to display in the Jupyter Notebook. Demo audience can change and add new SQL statements freely to, say, select a subset of the dataset, or split the dataset into training and test data. (2) We train an XGBoost model using the SELECT ... TO TRAIN syntax extension.  The audience can train any other models in the model zoo. (3) Using the SELECT ... TO PREDICT syntax extension, we make a prediction using the trained model. And (4) Using the SELECT ... TO EXPLAIN extension, we explain the model visually as an image showing in Jupyter Notebook.  The audience can change visual explainer and get other types of visualizations.

\paragraph{Episode Two.}  This episode shows SQLFlow works with Apache Hive and TensorFlow using the Iris Dataset\footnote{\url{https://en.wikipedia.org/wiki/Iris_flower_data_set}}.  This episode follows the same steps as Episode One but focusing on deep learning models other than tree models.

\paragraph{Episode Three.}  Differing from the previous two, this episode is about unsupervised learning.  It uses Alibaba MaxCompute and scikit-learn.  The dataset is the Electric Power Consumption\footnote{\url{https://www.kaggle.com/uciml/electric-power-consumption-data-set}}.  SHAP doesn't support explaining an unsupervised model, so we created our own tool with metaplot.  Episode Three consists of the same steps as the previous episodes.

\paragraph{Episode Four.} The episode shows how to use SQLFlow as a command line tool that compiles a SQL program to workflow, as explained in Section~\ref{sec:compile}.
This episode includes the following steps. (1) In the terminal, we type a command to convert the SQL program used in episode one into a Couler program. (2) Using the command-line tool couler, we convert the Couler program into a YAML file. (3) Using the workflow engine Argo's client tool, we submit the YAML for execution on a Kubernetes cluster. The audience is free to compare the SQL, Couler, and YAML files; they would notice that the SQL program has a few lines of code, the Couler program is longer than the SQL program, and the YAML file is very lengthy. In the dashboard of Argo in a Web browser, the audience is free to explore the visualized steps and their running progress. They are can also type Argo commands to monitor the progress and logs using command-line tools in the terminal. In the terminal, the audience can see the visual model explanation in the form of ASCII-art generated by a text-based backend we invented to work with SHAP.  From these steps, the audience might feel that sqlflow works like GCC as a compiler and couler works like LD as a linker as it generates an executable (YAML) file.

\bibliographystyle{ACM-Reference-Format}
\bibliography{sqlflow-sigmod}

\end{document}